# Separating active irradiation from passive outcome learning in radiotherapy: a finite-model computational benchmark

Yuntao Wang

Independent Researcher

Correspondence: yuntaowang00@gmail.com

ORCID: 0009-0005-5499-6258

## Abstract

**Background:** Informative irradiation may appear beneficial because of treatment reoptimization or improved use of outcomes obtained during static treatment, rather than the irradiation probe itself.

**Purpose:** To test whether gains from changing irradiation persist against optimized static treatment, passive outcome learning, and data-dependent stopping.

**Methods:** Three synthetic three-dimensional geometries supplied 51 parent fluence vectors. Four known response scenarios generated Bernoulli risks from normalized organ-dose contrasts at uncalibrated slopes of 0, 4, 16, and 64. A single unknown scenario applied to all 2,816 hypothetical uses. Total loss included acquisition and deployment. Net-value probe generation searched the parent-plan convex hull under two-operator risk constraints. Fixed passive and selected fixed active sampling were compared with passive optimal stopping and a restricted one-checkpoint, one-switch active policy. All methods shared a terminal treatment menu. Acquisition-menu ablations and frozen-policy sensitivity analyses were performed.

**Results:** Fixed active sampling improved on fixed passive sampling in three of nine nonzero-slope conditions, but never on passive stopping. At slope 64, passive stopping avoided 32.7811,

33.6779, and 17.8761 expected model events versus 29.2698, 30.3870, and 13.4817 for selected fixed active designs. One-switch adaptation added at most $4.00625 \times 10^{-5}$ expected events avoided. With the policies held fixed, two of four increments above the numerical reporting threshold reversed sign under alternative integration; all four became negative at some allowed prior weighting. Adding three generated probes changed none of 72 paired policy values, and none was used on a positive-probability path.

**Conclusions:** The tested probes did not demonstrate stable incremental value beyond optimized passive stopping. Treatment opportunities, stopping permissions, acquisition costs, and input sensitivity must be separated when attributing benefits to active irradiation. The results concern a restricted computational benchmark, not unrestricted active learning or clinical safety and effectiveness.

## Introduction

Radiotherapy planning must balance target coverage against exposure of normal tissues. These competing objectives mean that the value assigned to a treatment depends both on the achievable dose distributions and on how those distributions are expected to affect outcomes. Multiobjective planning and Pareto navigation provide established ways to represent such tradeoffs and to select among technically attainable treatments.[1,2] Any proposed learning strategy therefore has two potential sources of improvement: it may obtain useful information, or it may simply gain access to a better treatment than the original comparator.

Feedback-guided adaptation is not a new concept in radiotherapy. Adaptive radiation therapy was formulated as a process in which measurements update treatment planning during a course.[3] More recent work has considered biological information, treatment efficacy, and stopping decisions.[4] Mid-treatment imaging has also been used in a framework that permits either dose scaling or fluence-map reoptimization under response uncertainty.[5] These approaches establish that treatment and information can be connected, but they do not imply that deliberately changing irradiation to obtain information is preferable to using measurements generated during an already optimized treatment.

The general problem belongs to the exploration-exploitation tradeoff: an action can affect the immediate outcome and the information available for future choices. Dual-control formulations explicitly account for these two consequences.[6] In radiotherapy-motivated Bayesian uncertainty-directed dose finding, information criteria are combined with treatment utility and avoidance of undesirable doses.[7] A recent simulation study also integrates biomarker-derived response assessment, dose adaptation, and Bayesian trial design.[8] The present study neither introduces these general ideas nor evaluates the clinical effectiveness of any of those methods. It examines a narrower attribution problem in a shared, finite decision model.

Information must be valued against a decision and a time horizon. Expected-value-of-sample-information calculations address the consequences of learning before adoption, while decision-theoretic trial design relates that value to the cost of acquiring observations.[9,10] Sequential experimentation additionally allows observations to determine when acquisition should stop, including the possibility of not starting an experiment.[11] A method compared only with a fixed, poorly chosen observation count can therefore appear attractive even when its advantage disappears after optimizing the comparator's timing.

The observation plan and the eventual treatment plan need not coincide. Transductive experimental design explicitly distinguishes measurement actions from final recommendations.[12] Nevertheless, a measurement that separates response models is not necessarily the measurement that most improves a treatment decision. Model-discrimination design has its own established objectives.[13] If the separated models lead to similar treatment choices, or if creating the contrast incurs substantial outcome loss, improved discrimination need not translate into positive acquisition-inclusive value. This motivates optimizing the complete treatment-and-learning strategy rather than a separation metric alone. Controlled sensing also studies action-dependent observations for discriminating among known hypotheses, but its hypothesis-testing objectives differ from the acquisition-inclusive treatment loss considered here.[14]

Several other distinctions are important. Safe exploration requires explicit assumptions and constraints on what happens during learning; it does not follow automatically from favorable prior-average performance.[15] Identifying an optimal or approximately optimal action with specified error control is also different from minimizing cumulative expected loss.[16,17] Finally, estimating an unknown dose-response relationship from observational data is not the same task as selecting among supplied response laws. Causal effect estimation for high-dimensional radiotherapy exposure has been evaluated with synthetic dose data,[18] whereas the present benchmark supplies the candidate laws and leaves only their common scenario label unknown.

These distinctions suggest a practical test for an informative-irradiation proposal. Static treatment should first be reoptimized using the same knowledge and treatment opportunities. Outcomes obtained while applying that static treatment should then be allowed to influence later decisions. Fixed sampling and data-dependent stopping should be separated, and the marginal contribution of the newly generated acquisition plans should be tested while the terminal treatment menu remains identical. These comparisons need not be interchangeable or form a universal ranking; each isolates a different permission granted to a strategy. Retaining relevant treatment alternatives during both experimental design and final recommendation is also an established requirement of value-of-information analysis.[19]

We report a computational benchmark that implements these comparisons using plan-derived risks from three synthetic three-dimensional geometries. The primary question is whether a restricted outcome-dependent active strategy adds stable value beyond passive optimal stopping. Secondary questions concern fixed-sampling comparisons, direct net-value probe generation, and the acquisition-menu ablation. The benchmark is intended to expose attribution and verification requirements, not to establish a new radiation mechanism, an unrestricted optimal controller, or a calibrated clinical benefit. No clinical dataset is used to determine its risk parameters. Its contribution is the connected, reproducible comparison of these mechanisms and the limits of the resulting claims.

## Materials and methods

### Study design and scope

This exploratory computational study combines treatment-first plan construction, direct net-value probe generation, and restricted adaptive stopping in a shared finite-model benchmark. Some comparator refinements and diagnostics were informed by exploratory results; the study was not prospectively registered. The geometry-slope conditions share plans and response scenarios and do not constitute independent replications.

All observations in the decision model were Bernoulli outcomes evaluated by finite probability sums or dynamic programming. No patients, clinical images, animal experiments, biological measurements, or sampled outcome trajectories were used. The unit of time was one hypothetical treatment use, not a treatment fraction or calendar interval. Literature on clinical biomarkers informed interpretation but supplied no probabilities to the risk matrices. No independently validated radiation-transport calculation or clinical dose-response calibration was used.

The reporting separates the scientific aim, data-generating assumptions, target comparison, algorithms, and performance measures. This follows the organizational logic of the ADEMP framework for computational method evaluation,[20] while differing from a Monte Carlo simulation study: the reported policy values are evaluated by enumeration or recursion rather than estimated from repeated sampled trajectories. The geometry-slope conditions are designed cases, not a random sample from a clinical population. They are retained as separate results rather than treated as independent experimental replicates.

**Synthetic planning inputs and feasibility**

Three geometries were used: an offset uniform-density ellipsoid, an oblique low-density-insert geometry, and an asymmetric mixed-density geometry. Each contained one target and three organs at risk (OARs). Eight beam directions, each with a 5 by 5 beamlet array, provided 200 nonnegative fluence variables. A linear dose-influence operator combined primary attenuation, an empirical buildup factor, inverse-square scaling, and a Gaussian-blurred rectangular aperture. It was an analytical surrogate, without full scatter or secondary-electron transport, clinical source commissioning, absorbed-dose calibration, or multileaf-collimator sequencing.

The parent-plan construction used a 2-mm volume grid, additional 1-mm target points, sampled surfaces, and iterative constraint addition. Normalized target-dose bounds during optimization were 0.91-1.09 and the body maximum was 1.18; evaluation bounds remained 0.90-1.10 and 1.20. Target contrast, OAR spatial contrast, organ-mean increments, body-mean increments, fluence bounds, and adjacent-beamlet limits were also imposed. Geometry parameters and the executed constraint settings are provided in the Appendix section "Synthetic geometries and analytical dose operator"; these are mathematical settings, not prescriptions.

There were 16, 19, and 16 numerically distinct parent fluences, respectively. Treatment-oriented objectives comprised the four response endpoints, six pair averages, their equal-weight combination, and further static-optimization refinement. Separation-oriented probes were also included as parents. Thus the parent-only control was not restricted to conventional treatment endpoints; it excluded only the three newly generated net-value probes. This distinction matters when interpreting the probe ablation.

The archived parent assessment evaluated frozen weights on a shifted 1-mm volume grid and independently sampled boundaries under three design scenarios: all 306 plan-scenario-point-set checks passed. A second, out-of-design parameter stress scenario produced target undercoverage for every parent plan. These are finite-point results within the same surrogate, not continuous-space or independent physical certificates. They do not replace independently commissioned calculation-based dose verification.[21] The full computational reconstruction regenerated these fluences, score operators, and checks from the original geometry and configuration files; this is computational replay within the same surrogate, not independent physical validation.

The distinction between uncertainty represented during optimization and uncertainty encountered during evaluation is essential. Robust radiotherapy planning has long incorporated explicit uncertainty sets, including range uncertainty in proton planning.[22] Here the implemented sets are only parameter scenarios of an analytical surrogate. The existence of constraints under those scenarios does not justify borrowing physical accuracy or biological safety from the broader robust-planning literature. Surface points are used for spatial constraints, whereas organ averages use volume integration; the two roles are not combined into a single weighted mean.

**Response model and information available to the algorithms**

Let $w_0$ denote the fixed physical-reference fluence, $m_o^{(r)}(w)$ the mean surrogate dose in OAR $o$ under integration operator $r$, and $m_o^{(0)}(w_0) > 0$ its original normalization. For response scenario $f = 1, \ldots, 4$, define the dimensionless score

$$x_f^{(r)}(w) = \sum_{o=1}^{3} a_{fo} \frac{m_o^{(r)}(w) - m_o^{(r)}(w_0)}{m_o^{(0)}(w_0)}. \tag{1}$$

The four organ-weight vectors $a_f$ were (0.7, 0.2, 0.1), (0.1, 0.7, 0.2), (0.2, 0.1, 0.7), and (1/3, 1/3, 1/3). The fourth weight vector is the arithmetic mean of the first three. Its score is therefore their arithmetic mean for every fluence and integration operator, but its event probability is not generally their mean. Each integration operator induces the event-probability model

$$p_f^{(r)}(w) = \text{logistic}\left[\text{logit}(0.30) + \beta\, x_f^{(r)}(w)\right]. \tag{2}$$

Here $\text{logistic}(z) = (1 + e^{-z})^{-1}$ and $\text{logit}(p) = \log[p/(1-p)]$. Unless an operator is shown explicitly, $p_f(w) = p_f^{(0)}(w)$ denotes the nominal probability. The prior placed probability 1/4 on each scenario, and $\beta$ took values 0, 4, 16, and 64. The 30% probability belongs to the original reference $w_0$, not necessarily to the subsequently optimized static plan. Slopes, organ weights, and the reference probability were uncalibrated sensitivity settings. The scalar event was a constructed composite, not three separately validated organ complications, and $\beta$ had no per-Gy interpretation.

A single unknown scenario applied to all H = 2816 uses. Conditional on that scenario and the assigned plan, outcomes were independent, fully observed, and immediately available.

Algorithms knew all four plan-risk vectors and the prior, but not the realized scenario. Outcomes could therefore update the merits of plans that had not themselves been sampled. This is structural inference under a specified model, not independent causal validation of each treatment. Individual response heterogeneity, unmodeled scenarios, delayed toxicity, competing risks, tumor-control outcomes, and monetary measurement costs were not included.

**Treatment optimization and comparison structure**

For each geometry and β, the static anchor s minimized equal-prior average risk over the convex hull of the parent fluences. Because all parent logits were on the convex portion of the logistic curve, the objective was convex on that checked hull. A supporting-gradient bound assessed residual static suboptimality. This is a hull-specific numerical check, not a solution of the full 200-variable planning polytope.

Every retained generated probe was added to a common finite terminal menu. All strategies could select any plan in that menu when committing to the remaining uses. The static optimizer had access to the full parent hull containing every new probe, so it was not denied treatment opportunities granted to the learning strategies. Terminal decisions used the finite menu rather than solving a fresh continuous optimization at every posterior. Menu sizes after probe generation were 16, 21, and 17.

Table 1. Capabilities of the compared strategies.

| Strategy | Observation plan | Stopping permission | Optimization scope |
| --- | --- | --- | --- |
| Static treatment | Optimized anchor | No data-dependent change | Parent-hull mean-risk minimum |
| Fixed passive | Static anchor | Fixed sample size | All integer n from 0 to H |
| Selected fixed active | One selected hull plan | Fixed sample size | Local mixture search; integer refinement |
| Passive stopping | Static anchor | After each observation | All states for the fixed-anchor stopping problem |
| One-switch active | Static prefix, then one eligible probe | Prefix stopping; one checkpoint; probe stopping | At most one switch at the specified checkpoint |

All strategies share the same finite terminal treatment menu. Active and passive refer to acquisition, not whether terminal treatment can change. H = 2816 hypothetical uses. None of the active searches is claimed to solve unrestricted control.

The fixed-active and passive-stopping comparisons deliberately expose the effect of different timing permissions; they do not place fixed active sampling on an equal footing with unrestricted active control. The one-switch analysis then matches a limited form of outcome-dependent acquisition and stopping. Its active class contains the passive policy, so a nonnegative nominal increment follows from set inclusion. Strict magnitude, stability, and use of the new probes must be assessed separately.

**Acquisition-inclusive loss and gain**

For plan $j$ and response scenario $f$, write $P_{jf} = p_f(w_j)$ and $P_{sf} = p_f(s)$ for the static anchor. The prior satisfies $q_f \geq 0$ and $\sum_{f=1}^{4} q_f = 1$. For policy $\pi$ and binary outcome $Y_t$, define full-horizon gain by

$$G(\pi; q) = H \sum_{f=1}^{4} q_f P_{sf} - \mathrm{E}_{\pi,q}[\sum_{t=1}^{H} Y_t]. \tag{3}$$

Positive gain means fewer expected model events than using s for all H uses. Observation and deployment phases have the same event accounting. A switch probability is unconditional across complete experiments; an average number of probe uses is not the sample size conditional on switching. An event-probability difference is multiplied by 100 to express percentage points, whereas G is an expected count over the entire horizon.

A passive observation has zero immediate excess risk relative to s but consumes a use that could otherwise receive a different plan. Consequently, its opportunity cost remains in the finite-horizon decision problem. Physical delivery, data collection, waiting, and financial costs are not thereby zero in practice; they are outside this model. The principal comparison is the one-switch increment $G_{\text{active}}$ - $G_{\text{passive}}$ for the same anchor, prior, terminal menu, and evaluation operator. The acquisition-menu ablation separately tests whether the three newly generated probes alter this increment. Fixed-sampling comparisons are secondary benchmarks with less flexible stopping permissions.

**Direct net-value probe search and fixed sampling**

For a fixed acquisition plan *w* and count *n*, let $L_f(k;w)$ denote the complete binomial probability mass, including the combinatorial coefficient. Explicitly,

$L_f(k;w)=\binom{n}{k}p_f(w)^k[1-p_f(w)]^{n-k}$ for $k=0,\dots,n$. The dependence of $L_f$ on the fixed *n* is suppressed. The remaining $H-n$ uses receive the plan minimizing posterior mean event risk. Expected total loss is

$$J(w,n) = n \sum_{f=1}^{4} q_f \, p_f(w)$$

$$+ (H-n) \sum_{k=0}^{n} \min_j \sum_{f=1}^{4} q_f \, L_f(k;w) \, P_{jf} . \qquad (4)$$

A probe was a nonnegative mixture of parent fluences. Piecewise analytic gradients and sequential least-squares programming were used at n in {32, 128, 256, 512, 1024}, with seven prescribed starts per target and two menu-expansion passes. Across three geometries and three nonzero slopes, this gave 90 targets and 630 starts. Six starts did not report convergence. Accepted losses were ranked without an explicit solver-success filter; all 90 selected winners nevertheless reported convergence. This is multistart local optimization, not a global optimum over mixtures and sample sizes.

Each probe had to satisfy $p_f^{(r)}(w) - p_f^{(r)}(s) \leq 0.0009$ separately for every scenario under both the original 2-mm operator and the reused shifted 1-mm operator. The evaluation allowance was 0.001, or 0.1 percentage points. Both operators were design inputs at this stage. The condition constrains modeled observation-phase risk relative to s, not final deployment, all possible biological responses, or clinical safety. Full target records were retained, including repeated solutions.

After the terminal menu was closed, fixed passive sampling was optimized over every integer n from 0 to H. The same integer range was searched for the active candidate selected by the finite grid search in each nonzero condition. The active search could return s. Integer refinement was conditional on that candidate and did not globally reoptimize its fluence at every n.

**Passive optimal stopping**

After $n$ observations of the static plan $s$ and $k$ events, the posterior $\rho_f(n,k)$ is given in Equation (A7). Its predictive event probability is $r_{nk}=\sum_{f=1}^{4}\rho_f(n,k)P_{sf}$. With $H-n$ uses remaining, stopping commits to the minimum-posterior-risk terminal plan and has relative loss

$$S(n,k)=(H-n)\min_{j}\sum_{f=1}^{4}\rho_f(n,k)(P_{jf}-P_{sf}). \tag{5}$$

Writing C(n,k) for continuation loss, the optimal passive value follows the finite backward recursion

$$V(n,k)=\min(S(n,k),C(n,k)),$$

$$C(n,k)=(1-r_{nk})V(n+1,k)+r_{nk}V(n+1,k+1). \tag{6}$$

The recursion applies for $0\le k\le n<H$, with boundary condition $V(H,k)=0$. Immediate expected excess loss is zero because the acquisition plan is $s$; the use consumed by observation is accounted for by the decreasing horizon. This solves stopping for the fixed static acquisition plan and declared terminal menu, not richer passive strategies with external measurements, repeated deployment revisions, or additional acquisition plans. Independent absolute world-conditional evaluation and forward probability propagation were used to check the backward value.

**One-checkpoint active switching and probe ablation**

The adaptive strategy could stop during its initial static phase. At checkpoint T in {32, 128, 512}, it could retain the complete passive continuation rule or select one admissible acquisition plan using the observed count. After switching, it could stop after each outcome, with at most B in {64, 256} further observations. A second probe switch was not allowed. The checkpoint value was propagated backward through the prefix so that early stopping remained available.

Two acquisition menus were paired: parent plans plus the static anchor, and the augmented menu including the three new probes. Admissibility applied the 0.0009 modeled-risk guard under the original 2-mm and reused shifted 1-mm operators, with a probability tolerance of $10^{-10}$. It depended on supplied model inputs, not realized outcomes. The order-24 and order-40 values were used for frozen-policy sensitivity evaluation, not to determine eligibility. Both variants retained the identical full terminal menu. Three geometries, four slopes, three checkpoints, two caps, and two menus yielded 144 stored policies and 72 paired comparisons. Detailed recursions and tie handling are given in the Appendix section "Dynamic programs and policy inclusion".

**Input sensitivity and numerical verification**

After weights and decision rules were frozen, ellipsoidal product-integration operators of orders 24 and 40 recalculated organ means. These operators were available during adaptive-policy development and were not independent holdout inputs. Both observation probabilities and terminal losses were recomputed without reoptimizing policies or changing fluence normalization. They are not independent dose physics or rigorously bounded integration errors.

For fixed policies, the prior-weighted increment is linear in q. Sensitivity was assessed over $0.1 \leq q_f \leq 0.4$ with sum $q_f = 1$, whose six vertices each have two entries 0.4 and two entries 0.1. This range is a mathematical stress set, not a confidence region or calibrated distribution. The reporting threshold was $10^{-8}$ expected events, which is neither a significance level nor a clinically meaningful effect threshold. No Monte Carlo P values were calculated.

The primary performance quantity was the active-minus-passive gain within each geometry and slope, with identical prior, reference anchor, terminal choices, and evaluation operator. The ablation compared policies with and without access to the three new acquisition vectors, not policies with different terminal treatments. Activation probability and the mean number of nonstatic uses described how a rule obtained its value. All zero-slope and below-threshold conditions were retained. No multiplicity-adjusted significance claim was made because the engineering threshold only controls numerical reporting, not inference from sampled clinical data. Post hoc descriptive checks extracted static-risk vectors, conditional acquisition counts, and new-probe eligibility and checkpoint values from the frozen archive; they did not reoptimize plans or policies.

Verification distinguished file integrity, algorithm checks, and input stability. The archived modules contain independent scenario-conditional recursions, small history-tree checks, state-flow linear programs, and integration diagnostics. During full computational reconstruction, 52 parent-plan tests, 33 probe-generation tests, 26 adaptive-policy tests, and eight synthesis tests passed, and the independent absolute-risk evaluator reevaluated all 144 saved policies. The parent-dose matrices and plans, probe searches and integration inputs, and all adaptive policies were rebuilt using unchanged scientific configurations and newly generated upstream inputs. The appendix sections on numerical verification and reproduction distinguish numerical replay from physical or biological validation.

Verification was classified by what it can establish. Hash comparison detects changes to retained files; tests assess selected implementation invariants; an independently formulated recursion checks the evaluated policy loss; and alternative integration checks sensitivity to a numerical input. These activities do not validate the model for clinical use. The terminology follows the distinction between model verification and validation for an intended application.[23] In particular, agreement between two algorithms sharing the same risk matrix cannot demonstrate that the risk matrix represents patients.

All reported policy expectations are deterministic evaluations of a finite model. They do not have Monte Carlo sampling error, and no clinical population sample supports hypothesis tests or confidence intervals for these expectations. The three geometries were not sampled from a patient population. Consequently, comparisons are described by absolute increments, complete condition-level results, and sensitivity analyses rather than statistical significance.

**Software and computational implementation**

The analyses were implemented in Python 3.13.5 with NumPy 2.3.5, SciPy 1.17.0, and Numba 0.65.1 where compiled recursions were used. Linear programs used SciPy's HiGHS interface; the probe search used sequential least-squares programming (SLSQP). The general numerical library and the dual-simplex work underlying the HiGHS ecosystem are described separately;[24,25] citing the software does not certify an individual optimization result. Full configurations, deterministic starting points, stopping rules, and numerical tolerances are retained with the source code. Equations and implementation details are provided in the Appendix sections on geometry, local optimization, dynamic programs, and sensitivity.

The study involved no human participants, identifiable clinical records, patient images, animals, or biological specimens. ChatGPT (OpenAI) assisted with literature retrieval, mathematical formulation, software development, tool-executed computations, numerical diagnostics, and manuscript drafting. The checks were computational replays and independently formulated numerical evaluations, not replication by an independent human research team. AI systems were not treated as authors.

## Results

### Probe generation and comparison with fixed sampling

The direct search added three distinct fluences: none in the offset uniform geometry, two in the oblique low-density geometry, and one in the asymmetric mixed geometry. Most target optimizations returned existing plans. Selected fixed active designs exceeded optimized fixed passive sampling in three of nine nonzero-slope conditions; none exceeded passive optimal stopping. All zero-slope controls stopped without collecting data.

At $\beta = 64$, the additional gains of selected fixed active sampling over fixed passive sampling were 0.018322, 0.002854, and zero expected events. Passive stopping exceeded those selected fixed active values by 3.511302, 3.290915, and 4.394433 events, respectively (Table 2 and Figure 1). The winning fixed acquisition plans were pre-existing plans, not the newly generated probes. The reported improvement is a strategy-level expected count over the complete horizon. It is not a reduction of that many percentage points, and dividing by the horizon would yield only a model-average risk difference, not an observed clinical effect. The selected fixed-active method is allowed to return the static anchor, so equality is a legitimate result rather than a missing active run. The result therefore does not establish a distinct contribution of the new generation procedure.

Table 2. All geometry-slope comparisons.

| ID | β | Fixed passive | Selected fixed active | Passive stopping | One-switch increment |
|---|---|---|---|---|---|
| U | 0 | 0 | 0 | 0 | 0 |
| U | 4 | 0.02399195 | 0.02399195 | 0.03999799 | 4.909961e-14 |
| U | 16 | 2.520023 | 2.520023 | 3.611997 | 0 |
| U | 64 | 29.25152 | 29.26984 | 32.78114 | 4.994732e-08 |
| L | 0 | 0 | 0 | 0 | 0 |
| L | 4 | 0.02533549 | 0.02533549 | 0.04269189 | 1.545986e-13 |
| L | 16 | 2.300142 | 2.30089 | 3.315015 | 0 |
| L | 64 | 30.38416 | 30.38702 | 33.67793 | 5.468507e-09 |
| M | 0 | 0 | 0 | 0 | 0 |
| M | 4 | 0.001257414 | 0.001257414 | 0.001701848 | 6.393673e-07 |
| M | 16 | 0.3327614 | 0.3327614 | 0.5175943 | 4.006245e-05 |
| M | 64 | 13.48165 | 13.48165 | 17.87608 | 6.225352e-07 |

U, offset uniform; L, oblique low-density; M, asymmetric mixed. Unit: expected model events over H = 2816. Values below the $10^{-8}$ reporting threshold are retained and not treated as positive findings. The selected fixed-active search may return the static anchor.

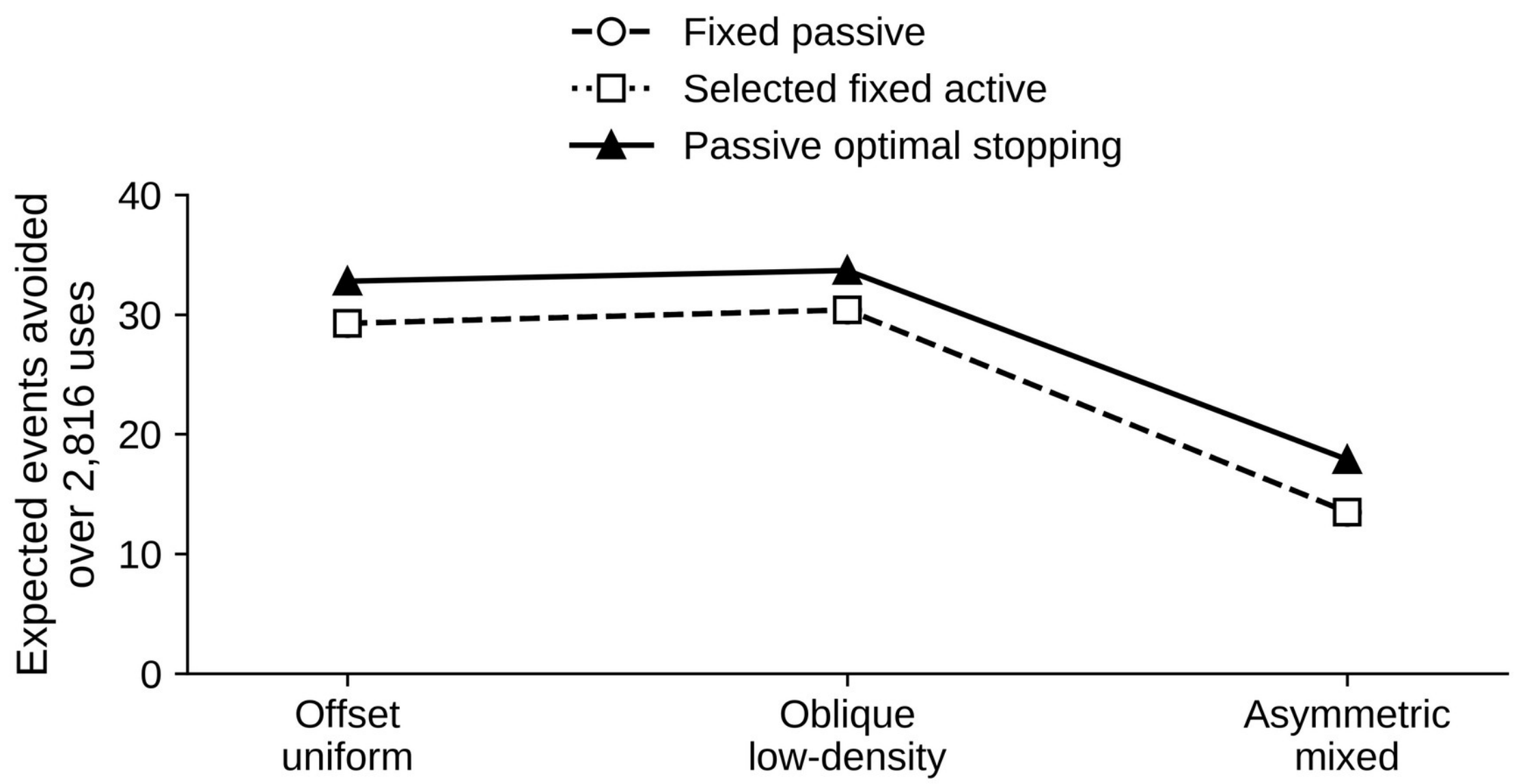

Figure 1. Acquisition-inclusive gains at the assumed response slope β = 64. All gains are relative to the same optimized static anchor within a geometry over H = 2816 hypothetical uses. Fixed-active designs use locally selected fluences and refined integer sample sizes. Passive stopping has more flexible timing permissions; the figure does not compare globally optimal active and passive control. Connecting lines are visual guides between categorical geometries, not interpolation over a continuous variable.

**Information available during static treatment**

All nine nonzero-slope conditions had four distinct static-anchor event probabilities (Table A4); all three zero-slope controls had probability 0.30 in every scenario. In each nonzero condition, the minimum-risk terminal endpoint differed across the four scenarios. For the asymmetric mixed geometry at slope 16, the static probabilities were 0.303733, 0.297286, 0.267051, and 0.289086. Thus static observations were informative about a scenario label that also affected treatment ranking. These descriptive properties permit passive learning, but do not separately quantify the effects of acquisition cost or restricted active control.

**One-switch adaptation and use of the new probes**

Four of nine nonzero-slope conditions had nominal increments above $10^{-8}$ events. The largest increment was $4.006245 \times 10^{-5}$ in the asymmetric mixed geometry at $\beta = 16$: passive gain was 0.5175942621 and one-switch gain was 0.5176343245. The retained checkpoint was $T = 512$ with $B = 256$. The probe was activated in 3.4129% of complete experiments, with an unconditional mean of 7.9891 nonstatic observations. Conditional on activation, the mean was 234.08 nonstatic observations out of the permitted maximum of 256; the unconditional mean does not describe the length of an activated episode.

All 72 parent-versus-augmented acquisition-menu pairs had equal values, and none of the three new probes was selected on a positive-probability path. The control already included prior separation-oriented probes and treatment plans. Thus the ablation shows no additional value from these three net-value-generated vectors within the tested architecture; it does not establish that all deliberately designed probes are unnecessary.

All three new probes were eligible at every tested slope in their own geometry: all 12 probe-slope pairs passed the two-operator guard with its stated numerical tolerance. Across the nine nonzero-slope pairs and six checkpoint-cap combinations, the 54 saved probe-value arrays represented 12,150 probe-specific checkpoint-state evaluations. No new probe improved the parent-menu continuation value by more than the implemented tie tolerance. Their nonuse was therefore not caused by exclusion from acquisition in these tests; it does not establish that a different generator or strategy class would have no value.

**Sensitivity of nominally positive increments**

Table A3 and Figure 2 report the four increments above the numerical reporting threshold. Two became negative under order-40 integration, including the largest, which changed to -0.0005744842 expected events. The other two retained positive signs. Because these evaluations use finite probability propagation, the changes are not Monte Carlo sampling fluctuations. They also do not establish which operator represents physical truth.

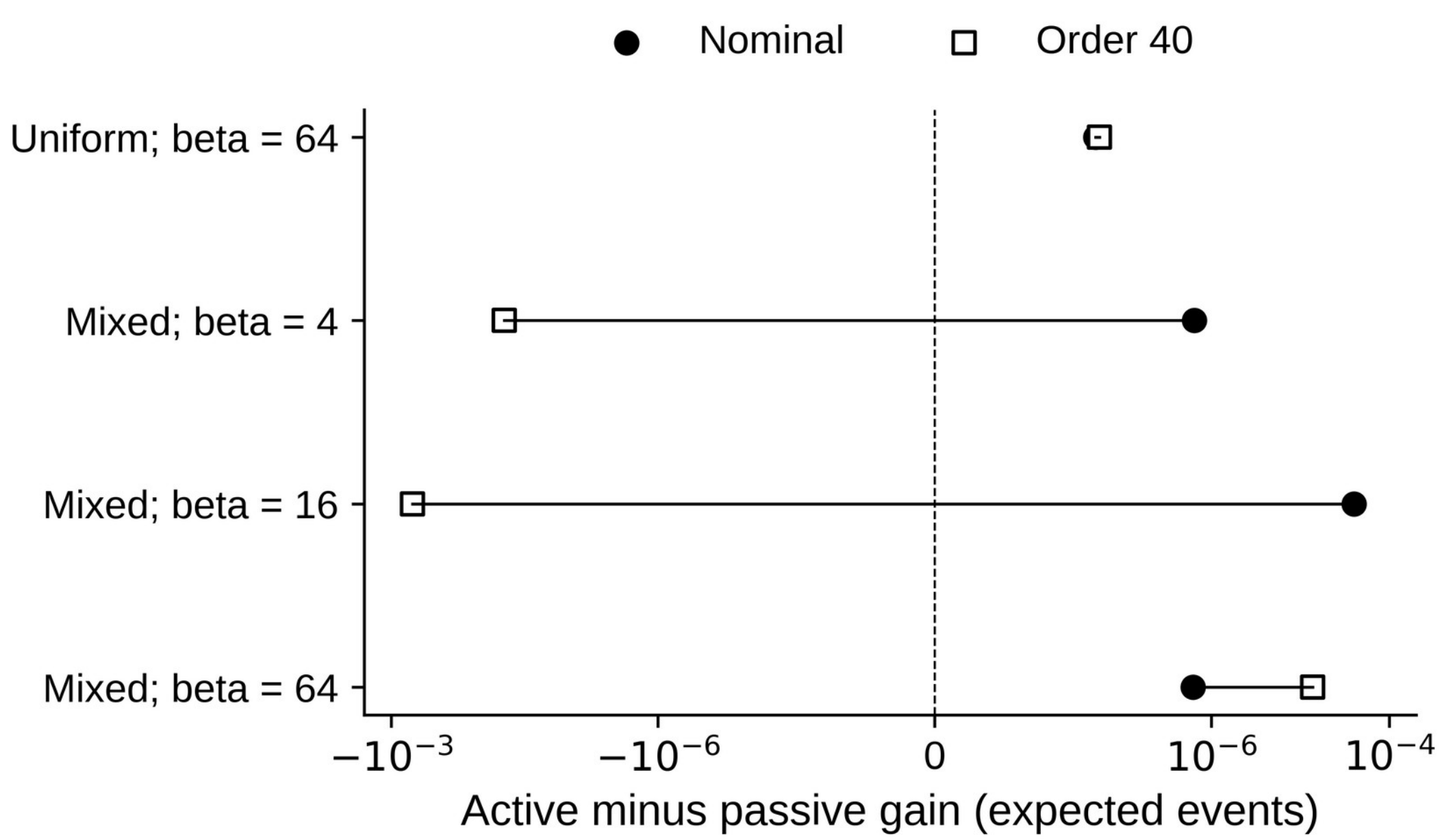

Figure 2. Frozen-policy one-switch increments under nominal and order-40 integration. The four conditions exceed the numerical display threshold of $10^{-8}$ expected events under the nominal operator. Both observation probabilities and terminal risks were reevaluated. The horizontal axis is symmetric-logarithmic with a linear interval from $-10^{-8}$ to $10^{-8}$. These are input-sensitivity comparisons, not confidence intervals.

All four frozen nominally positive policies had a negative increment at some allowed prior vertex (Figure 3). For the maximum nominal increment, the lowest prior-weighted value was

-0.0001143208. The four scenario-specific increments were approximately -0.000634459, +0.000297092, +0.000199973, and +0.000297643. A small positive prior average therefore coexisted with worse performance in one declared scenario. These are sensitivity calculations for fixed policies, not results of redesigning a robust policy.

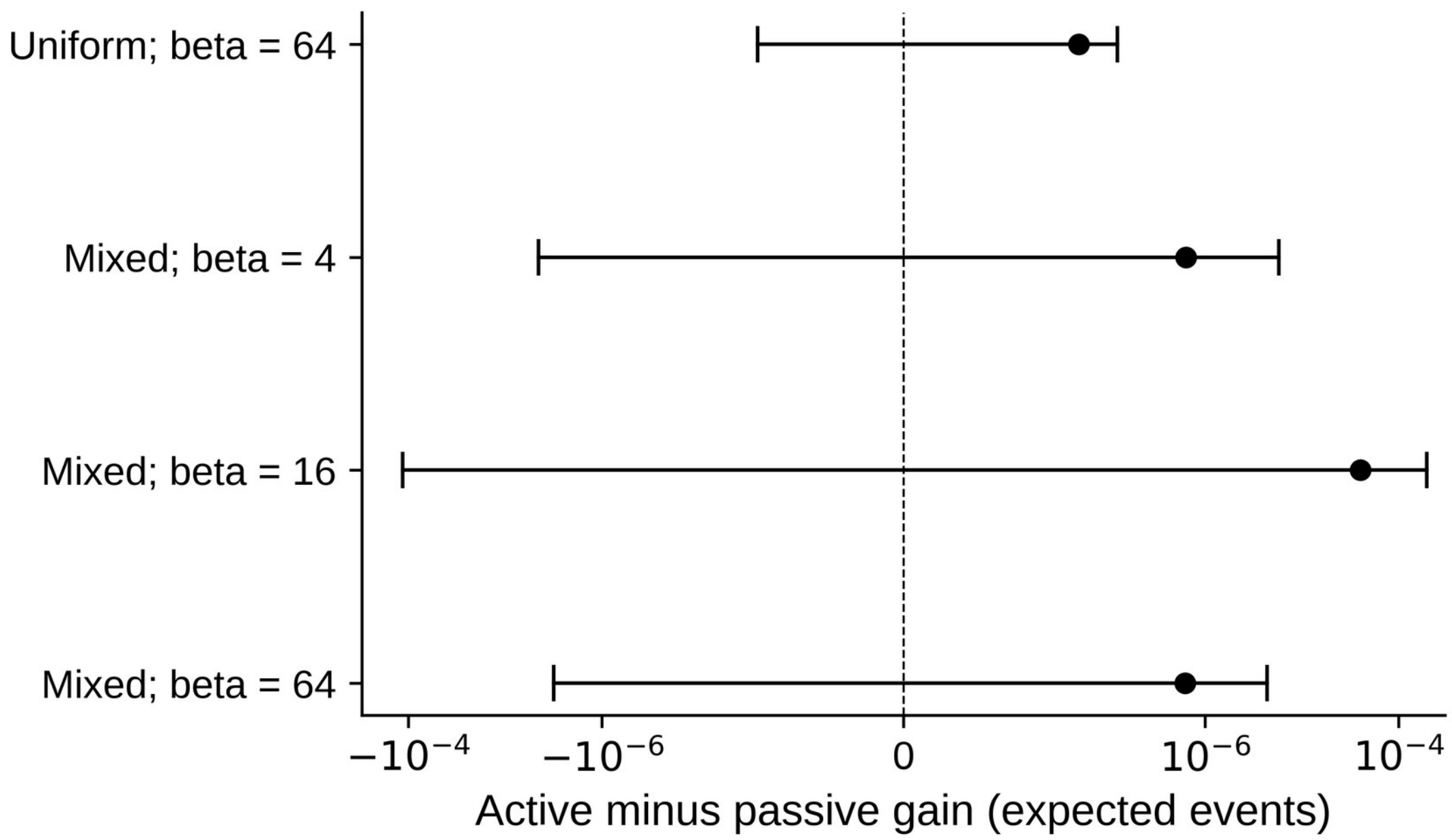

Figure 3. Sensitivity to prior weights for the same four frozen policy pairs. Segments show the smallest and largest increments over $0.1 \leq q_f \leq 0.4$ with sum $q_f = 1$; dots show the equal-prior value. The symmetric-logarithmic axis has a linear interval from $-10^{-8}$ to $10^{-8}$. The ranges are mathematical sensitivity sets, not statistical confidence intervals.

**Model-risk constraints and computational checks**

All 180 probe-target-by-integration evaluation records passed the 0.001 allowance. These records comprise 90 optimization targets assessed under order-24 and order-40 integration, not 180 distinct plans. The maximum evaluated modeled-risk increment was 0.000908130. These evaluation operators are distinct from the original 2-mm and reused shifted 1-mm operators used to impose the 0.0009 design guard. Among plans used in the adaptive probe phase, the largest order-40 increment was 0.000458746. No clinical or independent physical guarantee follows from these finite-model results.

The full computational reconstruction and result-verification suite passed 119 distinct tests. Independent absolute-risk evaluation of the 144 policies, representing 72 distinct saved rules, differed from the original scenario-specific gains by at most $6.77 \times 10^{-11}$ events; checkpoint minima agreed. The consolidated result file rebuilt from the newly generated results was byte-identical to the archived file; all 60 numeric cells in Tables 2 and A3 matched at their displayed precision. These checks support arithmetic consistency and resolution of the displayed increments. They do not establish that a tiny increment is important, that all possible implementation errors are excluded, or that the inputs are clinically credible.

## Discussion

### Principal findings and the contribution of the benchmark

The central finding is not that learning has no value. At the tested nonzero slopes, passive outcome learning improved expected performance over continued use of a static plan in the specified response families. Rather, most of the value available to the tested classes did not require the three newly generated irradiation probes. Optimizing passive stopping exceeded the selected fixed-active designs, and allowing a checkpoint-dependent active switch produced only very small nominal increments. Those increments were not uniformly stable under the retained integration or prior-weight changes.

This is a comparison benchmark rather than a new universal learning algorithm. Its contribution is that treatment opportunity, observation allocation, stopping flexibility, and use of newly generated fluences are evaluated within the same outcome accounting. The acquisition-menu ablation is especially informative: terminal options are held fixed, so an equal value cannot be attributed to denying a treatment to the control. Conversely, the ablation excludes only three additional vectors. Because the parent menu already contains separation-oriented probes, the result does not establish that purpose-built acquisition plans are unnecessary in general.

The observed ranking is also not a theorem ordering all passive and active methods. Passive stopping is exact within its fixed-acquisition, finite-menu problem; the active alternatives are locally optimized or structurally restricted. An unrestricted active controller includes passive behavior and could, in principle, do at least as well under the same correct model. The relevant negative result is that these implemented active extensions did not demonstrate a stable additional benefit, not that a more general optimization problem has no better solution.

**Relationship to previous treatment and learning methods**

The planning component relies on established multiobjective ideas. Pareto-surface construction and navigation make treatment tradeoffs explicit.[1,2] This matters because a historical plan need not be the best no-learning option under the biological assumptions subsequently supplied to an experimental method. Giving the experimental method a larger treatment menu or a different objective can produce improvement before any informative observation occurs. The common hull and terminal menu used here address that particular attribution problem; they do not establish that the hull contains the best physically deliverable treatment.

The information component has a similarly established history. Bayesian dose finding motivated by radiotherapy combines treatment utility with information-directed allocation.[7] Expected-value-of-information methods and decision-theoretic trial design assess whether evidence is worth its acquisition cost.[9,10] Delayed-response sequential models further show why a fixed sample size is not the only legitimate comparator.[11] Our findings complement these methods by quantifying how a probe's apparent advantage changes when the static comparator is allowed to use its own outcome information and to stop adaptively.

A distinct group of radiotherapy studies changes an individual patient's ongoing treatment using biological measurements. Optimal stopping in radiation therapy,[4] imaging-informed fluence adaptation,[5] and biomarker-informed Bayesian trial design[8] are close conceptual precedents. They do not solve the same benchmark: here each time step is a new hypothetical use under one common unknown scenario, not a fraction-specific decision in a patient with a measured biomarker. Therefore, the present results neither reproduce their reported advantages nor refute them. Likewise, they cannot establish novelty merely by adding the words adaptive, Bayesian, or informative to an irradiation strategy.

The benchmark is also different from recovering causal dose-response effects. Jenkins and colleagues investigated estimation for a high-dimensional radiotherapy exposure with a specified causal data-generating structure.[18] No comparable recovery problem is solved here: the four full response laws are known to the algorithms. The numerical exercises answer which policy to use under supplied laws, not whether those laws have been identified from clinical data. Keeping estimation, identification, and decision evaluation separate is necessary for interpreting the level of evidence.

### Why passive outcome learning can be strong

The distinct static probabilities and scenario-dependent terminal endpoints in Table A4 explain how passive observations can influence treatment choice in the specified model. Repeated outcomes from the static plan can therefore update scenario probabilities and change predictions for every plan linked through the common model. Passive does not mean uninformative, and stopping does not mean abandoning treatment: it means ending information acquisition and committing the remaining uses to a terminal plan. The opportunity cost of waiting is automatically represented because fewer uses remain after a long observation phase.

This information structure is also favorable to a narrowly specified model. A sequence of outcomes can distinguish among four known vectors much more readily than it can estimate an arbitrary dose-response surface. If two plausible laws agreed on the static plan but disagreed on treatment rankings elsewhere, observing only that plan would not resolve the difference. If a common near-optimal treatment existed across the laws, identifying the law might be unnecessary. These are logical alternatives, not newly executed experiments in this manuscript. They explain why conclusions depend on response support as well as prior probabilities.

Changing prior weights over the same four laws does not introduce either type of structural uncertainty. Nor does it model patients with independently varying latent response types. In a heterogeneous population, the information obtained from one patient can have a different value for the next patient or for the remainder of the first patient's treatment. The common-scenario calculation therefore cannot be relabeled as a validated individualized treatment system. Generalization would require a justified observation and response model rather than simply a larger number of runs of this one.

### Why improved discrimination need not improve complete treatment value

A probe can increase separation between predictions while worsening immediate outcomes or distinguishing models whose treatment choices are nearly identical. Classical model-discrimination design[13] and transductive measurement design[12] establish useful objectives, but those objectives are not interchangeable with acquisition-inclusive treatment value. In this study the direct probe objective includes both the observed uses and all subsequent uses, so a more informative probe must compensate for the costs of receiving it and for the delay before exploiting its information.

The distinction also clarifies why a more complex controller need not have a meaningful practical advantage. The one-switch class contains passive continuation, making its nominal optimum weakly better by construction. That inclusion result is not an empirical discovery. The questions that remain are the absolute increment, how often the added action is used, whether it depends on a particular new probe, and whether it persists under credible changes to the inputs. Here the answer was a very small increment, rare activation in the largest case, no marginal use of the three new probes, and nonuniform sensitivity.

The restricted search leaves legitimate possibilities unresolved. Posterior-dependent optimization outside the parent hull, multiple checkpoints, repeated switching, or other observation channels could add value. Equally, a more flexible passive comparator could use external measurements or allow repeated terminal revisions. These extensions would change the strategy permissions and must be evaluated as such. They are not grounds for asserting an unobserved advantage or for converting the current negative comparison into a universal impossibility claim. In addition, probes were generated using prior-weighted fixed-sampling loss, whereas the adaptive policy assessed them under checkpoint-specific posteriors with stopping. These objectives differ, so nonuse of the generated vectors does not establish an optimal probe generator for the adaptive decision problem.

**Numerical precision, decision objectives, and safety**

The largest nominal increment is larger than the reported discrepancies between the numerical evaluators, yet it is only $4.006245 \times 10^{-5}$ expected events over the full horizon. Its sign changed under the retained order-40 integration. These facts can coexist: implementation precision can be adequate while the scientific quantity is highly sensitive to the input operator. Neither printing additional decimal places nor repeating the same recursion resolves that sensitivity. The alternative operator is not physical truth, and the difference is not a confidence interval.

An expected-loss target also differs from fixed-confidence identification. Best-arm identification theory treats error guarantees and sample complexity explicitly,[16] while approximate-answer theory permits several sufficiently good recommendations.[17] Those guarantees cannot be borrowed for a posterior-mean terminal rule that was not designed to satisfy them. Conversely, a small probability of selecting the exact optimum need not imply large decision loss when several plans have similar risks. The present analysis reports its actual expected-loss objective without claiming a 95% selection guarantee or a clinically established minimum important difference.

Safety is yet another requirement. Favorable prior-average gain can coexist with a worse outcome in one declared scenario, as shown by the scenario-specific increments. The acquisition allowance constrains a model probability difference for a limited set of response laws, not realized adverse events or every patient's outcome. It does not separately protect tumor control, each organ endpoint, or the eventual terminal recommendation. Safe-exploration methods explicitly address such distinctions.[15] They should not be replaced by the assertion that a scalar expected objective improved.

**Physical validity and external relevance**

The geometries are synthetic instances from one construction, and the dose model is an analytical surrogate. Spatial checks and convex-mixture inheritance support a finite-set numerical feasibility claim. They do not establish absorbed dose, continuous-space extrema, full transport accuracy, or machine deliverability. The retained out-of-design undercoverage is therefore part of the evidence boundary, not an inconvenient result to remove. Explicit uncertainty-aware treatment optimization is well established,[22] but a declared uncertainty set only protects the effects it actually represents.

Clinical implementation would need commissioned beam and delivery data, appropriate end-to-end verification, and independent calculation checks such as those addressed by AAPM Task Group 219.[21] It would also need a biological endpoint and a credible dose-response relation. None is supplied by an agreement between two software evaluators. The separation of implementation verification from intended-use validation is central to model evaluation.[23] The study can support methodological conclusions without a clinical trial, provided that no clinical efficacy or safety claim is inferred.

Further restrictions include immediate complete outcomes, no monetary cost or calendar-time penalty, a fixed prior, and a finite terminal menu. The latter can omit the best posterior-specific convex mixture even though the initial static anchor was optimized over the hull. More permissive terminal optimization could improve both active and passive values, and its effect on their difference is not known. The comparison also has no calibrated connection between the composite event and particular patient-reported or tumor-control endpoints. The reported fixed and adaptive comparisons used only $H = 2816$; they do not establish stability to a different horizon.

### Reporting, interpretation, and next evidence requirements

The comparisons are exploratory rather than confirmatory. Complete condition-level results, zero-slope controls, and numerical sensitivity checks are therefore reported alongside nominal gains. Shared geometries, plans, and response laws limit the independence of the comparisons.

For future informative-irradiation evaluations, the comparator should receive the same treatment opportunities and stated model knowledge, while differences in timing and action permissions should be visible. Probe ablation should hold terminal choices fixed. Acquisition and deployment loss should be reported together, and positive increments should be checked under inputs not used to construct them. The ADEMP reporting structure offers useful organization,[20] but implementing a reporting framework cannot compensate for a response model without supporting evidence.

The next evidence requirement depends on the intended claim. A stronger numerical-method claim would require a less restricted active search, appropriate optimality bounds or strong baselines, and genuinely new model instances. A claim about treatment benefit would additionally require physically validated plans and a defensible response or measurement mechanism. The staged distinction between modeling and clinical evaluation is consistent with R-IDEAL.[26] For the present study, the justified conclusion remains narrower: these generated probes did not provide a stable additional benefit beyond the optimized passive comparator in the tested finite model.

## Conclusions

In three synthetic geometries with four specified response laws, passive optimal stopping outperformed the selected fixed active designs at every nonzero slope; all zero-slope controls tied. One-checkpoint adaptation added at most $4.00625 \times 10^{-5}$ expected events avoided over 2,816 uses, and the three newly generated probes changed no paired policy value. With the policies held fixed, the small nominal increments were not uniformly stable to available integration and prior-weight variations. This finite-model benchmark supports aligning treatment choices, stopping opportunities, and acquisition costs before attributing benefit to active irradiation. It does not establish the futility of unrestricted active learning or the clinical safety or effectiveness of any evaluated policy.

## Data and code availability

The source code, fixed configurations, synthetic planning inputs, derived dose-score and risk matrices, saved policies, numerical results, and verification records supporting this study are available from the corresponding author on request. The materials include the full reconstruction workflow and companion scripts and data for the descriptive diagnostics. No patient-level data are included. Their computational scope is described in the Appendix section "Computational reproducibility and materials".

## Ethical scope

The work used synthetic geometries, specified response laws, and computed outcome expectations. It did not enroll human participants or use identifiable clinical records, animal experiments, or biological specimens.

## Appendix. Detailed methods and reproducibility

### Computational scope and units

The computational workflow comprises parent-plan construction, net-value probe generation within the parent-plan convex hull, and evaluation of fixed and adaptive acquisition strategies. All components use the same geometries and specified response laws; they constitute one integrated comparison rather than independent validation studies. No clinical effect estimates were used to calibrate the synthetic risk matrices.

$H = 2816$ denotes a fixed number of hypothetical uses. A gain of 0.001 is one thousandth of an expected event across all H uses; it is not a 0.1% risk reduction. A probability contrast of 0.001 is 0.1 percentage points. The reporting threshold $10^{-8}$ expected events is only a numerical display rule. Conditional on a fixed response scenario, outcomes are independent across uses. No participant-level outcomes were sampled, and no individual latent-type model was fitted.

### Synthetic geometries and analytical dose operator

All geometries use a body ellipsoid with semiaxes (9, 8, 7) cm. Ellipsoidal targets and OARs are defined in a shared coordinate system; an ellipsoid includes points with the sum of squared normalized coordinates no greater than $1 + 2 \times 10^{-12}$. Target points take precedence, followed by OARs in their saved order, so OAR mean scores exclude overlap with previously assigned structures. Density insert coordinates affect attenuation, not the definition of a clinically characterized tissue.

Table A1. Geometric parameters of the parent-plan models.

| ID | Structure | Center (cm) | Semiaxes (cm) | Density |
|---|---|---|---|---|
| U | Target | 0.25, -0.2, 0.3 | 1.6, 1.55, 1.8 | - |
| U | OAR 1 | 3.05, 0.25, 0.35 | 1.05, 1.8, 1.9 | - |
| U | OAR 2 | -0.3, 3.15, -0.1 | 1.9, 1, 1.8 | - |
| U | OAR 3 | -3.15, -1.4, -0.45 | 1.25, 1.5, 1.9 | - |
| L | Target | -0.2, 0.25, -0.15 | 1.65, 1.45, 1.85 | - |
| L | OAR 1 | 3, -0.3, 0.6 | 1.05, 1.75, 1.9 | - |
| L | OAR 2 | 0.5, 3.2, -0.7 | 1.85, 1.05, 1.7 | - |
| L | OAR 3 | -3.15, -1.35, 0.45 | 1.2, 1.55, 1.85 | - |
| L | Insert 1 | -3.25, 0.35, 0.5 | 2.2, 2.8, 3 | 0.35 |
| M | Target | 0.15, 0.15, -0.25 | 1.7, 1.4, 1.8 | - |
| M | OAR 1 | 3.2, 0.2, -0.65 | 1, 1.8, 1.9 | - |
| M | OAR 2 | -0.4, 3.1, 0.75 | 1.8, 1, 1.7 | - |
| M | OAR 3 | -3, -1.7, 0.2 | 1.15, 1.5, 1.85 | - |
| M | Insert 1 | -3.4, -0.2, -0.1 | 2.3, 2.7, 3.2 | 0.35 |
| M | Insert 2 | 3.4, -2.15, 0.25 | 0.9, 1.1, 2.7 | 1.5 |

U, offset uniform; L, oblique low-density; M, asymmetric mixed; OAR, organ at risk. Body center: (0, 0, 0); semiaxes: (9, 8, 7) cm. Densities are relative numerical insert densities. A dash indicates that the structure does not override the background density.

There are eight gantry angles (0, 45, 90, 135, 180, 225, 270, 315 degrees), with elevations (0, 15, 0, -15, 0, 15, 0, -15 degrees). Each source is 100 cm from the target center. The 5 by 5 aperture-center grid in each beam has width and spacing 1.1 cm. Let l(v) be the path length through the body plus the sum of ($\rho_i$-1) times the path length through insert i. For each beam, the coded kernel is

$$K(v) = \exp[-\mu_0 c_\mu l(v)][1-\exp(-l(v)/b)]\left(\frac{100\,\mathrm{cm}}{r(v)}\right)^2 . \tag{A1}$$

Here $\mu_0 = 0.05\,\mathrm{cm}^{-1}$, $b = 0.5\,\mathrm{cm}$, $c_\mu$ is the scenario attenuation multiplier, and $r(v)$ is the source-to-point distance. All distances are evaluated in cm. The Gaussian aperture width is $\sigma(v) = c_{\sigma}[0.18\,\mathrm{cm} + 0.007\,l(v)]$. In each transverse beam coordinate $u$, a rectangular aperture centered at $u_i$ contributes

$$F_i(u,v) = \Phi\left(\frac{u_i + a/2 - u}{\sigma(v)}\right) - \Phi\left(\frac{u_i - a/2 - u}{\sigma(v)}\right). \quad \text{(A2)}$$

where a = 1.1 cm and Φ is the standard normal cumulative distribution. The beamlet influence is K(v) times the two transverse factors; dose is the nonnegative fluence-weighted sum of all 200 influences. These formulas define a mathematical surrogate, not energy-conserving coupled radiation transport or absorbed dose in Gy. Relative densities 0.35 and 1.5 describe numerical inserts and are not sufficient to model lung or bone composition.

Transverse aperture coordinates are projected along the source-to-point ray onto the plane 100 cm from the source. Specifically, the displacement component along each transverse unit vector is multiplied by 100 cm and divided by its component along the beam-axis unit vector. The source lies 100 cm upstream of the target center. This projection and the path-length calculation follow the archived analytical implementation; no additional transport physics is implied.

The five (attenuation, aperture-width) multiplier pairs were (1.00,1.00), (0.97,0.90), (1.03,1.10), (0.94,1.20), and (1.06,0.80). The first three entered planning; the last two were stress scenarios. Parent verification passed the former and the first stress scenario, whereas all 51 parents failed the second stress scenario. These retained failures exclude an unrestricted robustness claim.

The parent separation pool combined a 2-mm volume grid, 1-mm target interior points, and 2,048 points per sampled surface. Surface points constrained local dose but did not change the volume integral used for mean dose. In the parent-plan verification, the shifted 1-mm grid used offsets (0.023,0.047,0.083) cm and each evaluation surface used 8,192 newly oriented points.

Exact input definitions and masks are archived; full computational reconstruction regenerated the planning matrices and both validation point sets from these configurations.

Table A2. Planning guards and evaluation limits.

| Quantity | Planning value | Evaluation value |
|---|---|---|
| Target dose | 0.91 to 1.09 | 0.90 to 1.10 |
| Body maximum | 1.18 | 1.20 |
| Target absolute contrast | 0.018 | 0.020 |
| OAR spatial contrast | 0.09 x reference OAR mean | 0.10 x reference OAR mean |
| OAR mean increment | 0.003 x reference mean | 0.005 x reference mean |
| Body mean increment | 0.008 x reference mean | 0.010 x reference mean |
| Fluence bounds | 0 to 4 | Same bound |
| Adjacent beamlet difference | 1.5 | Same bound |

Dose is normalized, not expressed in Gy. Contrast bounds refer to the fixed original planning reference. The modeled risk guard imposed later is distinct from these point-dose and fluence constraints.

**Fixed reference, score operators, and plan mapping**

For geometry g, the archived reference $w_0$ minimizes a planning mean-dose objective and supplies positive nominal OAR normalizers $m_o^{(0)}(w_0)$. This reference is not automatically the later Bayesian static optimum. For any integration rule r, both w and the same fixed $w_0$ are evaluated with that rule, but the nominal denominator remains fixed. In vector form,

$$x_f^{(r)}(w) = \sum_{o=1}^{3} a_{fo} \frac{m_o^{(r)}(w) - m_o^{(r)}(w_0)}{m_o^{(0)}(w_0)}. \quad \text{(A3)}$$

No fluence renormalization, re-centering on a newly chosen plan, or fitting to the observed result is performed during sensitivity analysis. Let $X_{if}^{(r)}$ be the score of parent i. Linearity gives $x_f^{(r)}(\sum_i \lambda_i w_i) = \sum_i \lambda_i X_{if}^{(r)}$ whenever $\lambda_i \geq 0$ and $\sum_i \lambda_i = 1$. Probabilities are then obtained by the nonlinear logistic mapping, rather than by mixing parent probabilities.

The parent sets contain 16, 19, and 16 vectors. They include the original reference, scenario-specific treatment endpoints, pairwise objectives, static-refinement points when distinct, and separation-oriented probes. After net-value probe generation, terminal menus contain 16, 21, and 17 vectors, respectively. The three additional vectors occur only in the second and third geometries. The exact mapping from optimization target records to distinct menu vectors is stored in the probe-generation file plans.npz, including menu_names, menu_lambda, parent_W, and record_to_menu.

The 90 probe-optimization targets do not represent 90 new plans. Removing the three generated net-value vectors from acquisition leaves the separation-oriented probes in the parent menu. The ablation therefore tests the incremental contribution of those three vectors, not the value of all deliberately designed probes.

**Convex feasibility and the static bound**

If all parents satisfy a common linear system $C_r w_i \leq b_r$, then a nonnegative convex mixture satisfies it as well. Absolute contrast and adjacent-fluence bounds are represented by pairs of linear inequalities. The property applies only to the operator, points, scenarios, and normalization represented by that system. It provides no certificate for an unseen anatomy, continuous-space maximum, another physical engine, or a deliverable leaf sequence.

For mixtures $\lambda \in \Delta$, let $F(\lambda)=\sum_{f=1}^{4} q_f p_f(\sum_i \lambda_i w_i)$, where $\Delta$ is the simplex defined by $\lambda_i \geq 0$ and $\sum_i \lambda_i = 1$. The second derivative of the logistic function with respect to its scalar argument is $p(1-p)(1-2p)$. All parent logits in the checked input hull correspond to probabilities below one half. Convex mixing remains in that interval, so $F$ is convex on this hull. With $g=\nabla F(\lambda_s)$ at the returned static mixture, the first-order convexity inequality gives

$$0 \le F(\lambda_s) - \min_{\lambda \in \Delta} F(\lambda) \le g^{\mathrm{T}} \lambda_s - \min_i g_i . \tag{A4}$$

This provides an upper bound on the static objective's residual error inside the simplex. It is not a certificate of the best plan in the original continuous planning polytope. The archived statics and their tangent-gap checks are preserved. At β = 0, all risks equal 0.30 and information value is zero.

For $\beta > 0$, the guard $p_f^{(r)}(\lambda) \le p_f^{(r)}(s) + c$ can be transformed into a linear score constraint because logit is increasing. Its right side is $[\operatorname{logit}(p_f^{(r)}(s)+c) - \operatorname{logit}(0.30)]/\beta$, provided the logit argument is strictly between zero and one. The design allowance was $c=0.0009$ for all four scenarios under both design operators; the evaluation limit was 0.001. The guard does not restrict terminal deployment or unknown response laws. If the per-use guard holds exactly, at most $B$ probe observations imply an upper bound $Bc$ on their cumulative latent expected excess in a declared scenario, not on realized adverse-event counts. The implemented admissibility check allows a $10^{-10}$ probability tolerance; a numerical bound must include that tolerance rather than assert exact zero excess.

### Fixed-sample policy and local probe optimization

For fixed *w* and *n*, use the same binomial mass $L_f(k;w)$ as in Equation (4). A terminal action $j^*(k)$ minimizes $\sum_{f=1}^{4} q_f L_f(k;w) P_{jf}$. The common posterior-normalization factor and binomial coefficient cancel for action selection, but the coefficient must be retained when averaging over counts. Consequently,

$$J(w,n) = n\sum_{f=1}^{4} q_f\, p_f(w)$$

$$+(H-n)\sum_{k=0}^{n} \min_j \sum_{f=1}^{4} q_f\, L_f(k;w)\, P_{jf}. \quad \text{(A5)}$$

Gain is $H\sum_{f=1}^{4} q_f P_{sf} - J(w,n)$. The optimizer may return the static plan $s$, and $n=0$ represents no acquisition. For a fixed terminal menu and unchanged minimizing action on each count branch, write $p_f = p_f(w)$ and $X_{if} = x_f^{(0)}(w_i)$. Differentiation with respect to mixture coordinate $\lambda_i$ gives

$$\frac{\partial J}{\partial \lambda_i} = \beta \sum_{f=1}^{4} X_{if}\, q_f [n\, p_f(1-p_f) + (H-n) A_f],$$

$$A_f = \sum_{k=0}^{n} L_f(k;w)\, P_{j^*(k),f}(k - n\, p_f). \quad \text{(A6)}$$

At an action tie, this is a branch derivative of a piecewise-smooth function, not a guarantee of differentiability or global convergence. SLSQP used ftol = $10^{-10}$, at most 250 iterations, and seven starts per target: the static mixture, four cap-clipped scenario endpoints, and two deterministic Dirichlet mixtures. The seed was 2026100704. Two menu passes, five sample-grid values, three nonzero slopes, and three geometries gave 630 starts across 90 targets. Six starts did not report convergence. Accepted losses were ranked without an explicit solver-success filter; all 90 selected winners nevertheless reported convergence. Original failed starts remain in the archived generation records.

After menu closure, all n = 0,...,H were evaluated for passive acquisition at s. The same integer search was applied to the selected active candidate per nonzero condition. This is exhaustive in n only after fixing w. It is not exhaustive joint optimization over every w and n, and should not be labeled a globally optimal active design.

**Dynamic programs and policy inclusion**

For passive observations, the posterior after state (n,k) is

$$\rho_f(n,k)=\frac{q_f P_{sf}^k(1-P_{sf})^{n-k}}{\sum_{u=1}^{4} q_u P_{su}^k(1-P_{su})^{n-k}}. \tag{A7}$$

The binomial coefficient cancels in Equation (A7) because it is common to all four scenarios.

Using $r_{nk}=\sum_{f=1}^{4}\rho_f(n,k)P_{sf}$ and the stopping loss in Equation (5), the relative-loss recursion is

$$V(n,k)=\min(S(n,k),C(n,k)),$$

$$C(n,k)=(1-r_{nk})V(n+1,k)+r_{nk}V(n+1,k+1). \tag{A8}$$

The boundary condition is $V(H,k)=0$. Immediate expected excess loss is zero because acquisition uses the static plan $s$. Each continuation reduces the remaining horizon by one, thereby retaining the opportunity cost of delayed deployment. This recursion solves the fixed-acquisition stopping problem; it does not enforce confidence-interval coverage or a frequentist terminal-error guarantee.

At checkpoint $T$, let $k_0$ be the number of static-plan events. After switching to plan $a$, write $(m,k)$ for the number of additional observations and events. The posterior $\rho_{af}(m,k)$ is proportional to $q_f P_{sf}^{k_0}(1-P_{sf})^{T-k_0}P_{af}^k(1-P_{af})^{m-k}$. Dependence on the fixed checkpoint data $(T,k_0)$ is suppressed. Define $r_a(m,k)=\sum_{f=1}^{4}\rho_{af}(m,k)P_{af}$. Also, $c_a(m,k)=\sum_{f=1}^{4}\rho_{af}(m,k)(P_{af}-P_{sf})$. The former is the predictive event probability; the latter is immediate expected excess loss. With $S_a$ denoting deployment loss and $Q_a$ continuation loss,

$$U_a(m,k)=\min(S_a(m,k),Q_a(m,k)),$$

$$S_a(m,k)=(H-T-m)\min_j \sum_{f=1}^{4} \rho_{af}(m,k)(P_{jf}-P_{sf}),$$

$$Q_a(m,k)=c_a(m,k)+(1-r_a(m,k))U_a(m+1,k)$$

$$+r_a(m,k)U_a(m+1,k+1). \quad \text{(A9)}$$

Equation (A9) applies for $0 \le k \le m < B \le H-T$. At $m=B$, observation is no longer permitted and $U_a(B,k)=S_a(B,k)$. At the checkpoint, the rule chooses the smaller of $V(T,k_0)$ and the eligible values $U_a(0,0)$. These values are propagated backward through the static prefix, where early stopping remains available. Choosing passive continuation at the checkpoint does not create a later switching opportunity.

The adaptive-policy grid used T in {32,128,512} and B in {64,256}. Parent and augmented acquisition menus shared the same complete terminal menu. There were 144 stored policies and 72 menu pairs. Since the active checkpoint always includes the passive continuation, backward induction gives $V_{\text{active}} \le V_{\text{passive}}$, or $G_{\text{active}} \ge G_{\text{passive}}$, in the nominal model up to numerical tolerance. Nonnegative increments are therefore partly a design identity. The strict size of the increment and its sensitivity are empirical properties of the specified numerical instance.

The implementation uses log probabilities and a $10^{-12}$ expected-event tie tolerance. The relative-loss recursion avoids subtracting large absolute totals at each optimization step. Forward evaluation omits node masses below $10^{-250}$. A conservative finite-node bound multiplies this cutoff by the horizon and the number of potentially omitted nodes; this bound concerns discarded probability mass only. It does not bound floating-point arithmetic, dose integration, biological model error, or total scientific uncertainty. Terminal-action ties with no uses remaining have no value effect. Original code and fixture outputs retain implementation details.

**World-conditional policy evaluation**

Policy values were checked without re-solving the Bayesian optimization. Conditional on scenario f, at a stop node selecting j with h uses remaining, absolute future loss is h $P_{jf}$. At a node sampling a, it is $P_{af}$ plus (1-$P_{af}$) times the no-event successor loss and $P_{af}$ times the event successor loss. This direct recursion and forward probability propagation evaluate the complete frozen rule, including prefix observations, active observations, and deployment.

Acquisition and deployment counts must sum to H in every scenario in expectation. The switching probability is averaged over all complete experiments. The unconditional nonstatic count can be small because switching is rare; it must not be interpreted as the active-phase count among switched experiments. In the maximum-increment condition, activation was 0.0341290452 and unconditional nonstatic observations were 7.98905493. Their ratio gives 234.08375 nonstatic observations conditional on activation.

The gain conditional on scenario f is H $P_{sf}$ minus total evaluated absolute loss. Averaging with the design prior gives the reported value. For a frozen active/passive pair, subtracting these conditional gains yields $\Delta_f$; alternative prior evaluation is simply $\sum_f q_f \Delta_f$. This calculation does not fit a new rule under that prior.

**Integration and prior sensitivity**

The nominal design operator used volume-only 2-mm organ means. The shifted 1-mm operator was included in the probe-generation risk constraints and thus was no longer held out. Subsequent ellipsoidal integration used Gauss-Legendre rules in radius and cosine of polar angle, and 2n equally spaced azimuthal nodes shifted by 0.371 of a node interval, with n = 24 or 40. The product grids contain $2n^3$ points per ellipsoid before overlap masking. Radial weights include $3r^2$ and are renormalized over the retained organ mask. The original reference fluence and nominal mean-dose denominators remain fixed.

These two operators were applied after probe weights and fixed-sampling rules were frozen. They were already available during adaptive-policy development and were not independent holdouts. Frozen evaluation changes both transition probabilities and terminal risks. It is not adequate to change only final risk while retaining the old observation distribution. The resulting checks test available numerical-input sensitivity, not independent transport, a formal integration bound, or a clinical uncertainty distribution.

For prior sensitivity, $q_f$ belongs to [0.1,0.4] and sums to one. There are six vertices, with two entries 0.4 and two 0.1. For a fixed policy pair, the extrema of the linear function $\sum_f q_f \Delta_f$ occur at these vertices. The set was specified for sensitivity analysis, not estimated from patient data. A negative vertex value does not prove a policy redesigned for that prior would also fail.

Table A3. One-switch increments and frozen-policy sensitivity.

| **Geometry; β** | **Nominal** | **Order 40** | **Lowest prior value** |
|---|---|---|---|
| Offset uniform; 64 | 4.99473e-08 | 5.53349e-08 | -2.46431e-08 |
| Asymmetric mixed; 4 | 6.39367e-07 | -5.31727e-05 | -4.51184e-06 |
| Asymmetric mixed; 16 | 4.00625e-05 | -0.000574484 | -0.000114321 |
| Asymmetric mixed; 64 | 6.22535e-07 | 1.36433e-05 | -3.15298e-06 |

Unit: expected model events over H = 2816. Negative increments favor passive stopping. Only nominal increments above $10^{-8}$ are shown; all 12 conditions appear in Table 2. Prior extrema are sensitivity values, not confidence limits.

Table 2 retains all slope conditions, including controls and increments below the display threshold. Table A3 and Figures 2 and 3 separately display the four increments above the numerical reporting threshold. Complete unfiltered numerical results are retained in results/consolidated.json.

Table A4. Static-anchor event probabilities in the nominal response model.

| ID | β | Scenario 1 | Scenario 2 | Scenario 3 | Scenario 4 |
|---|---|---|---|---|---|
| U | 0 | 0.300000 | 0.300000 | 0.300000 | 0.300000 |
| U | 4 | 0.301318 | 0.300236 | 0.294580 | 0.298703 |
| U | 16 | 0.305290 | 0.300944 | 0.278668 | 0.294830 |
| U | 64 | 0.321435 | 0.303608 | 0.220743 | 0.279643 |
| L | 0 | 0.300000 | 0.300000 | 0.300000 | 0.300000 |
| L | 4 | 0.301467 | 0.300330 | 0.295542 | 0.299107 |
| L | 16 | 0.305894 | 0.301321 | 0.282404 | 0.296437 |
| L | 64 | 0.323955 | 0.305306 | 0.233548 | 0.285896 |
| M | 0 | 0.300000 | 0.300000 | 0.300000 | 0.300000 |
| M | 4 | 0.300931 | 0.299321 | 0.291546 | 0.297250 |
| M | 16 | 0.303733 | 0.297286 | 0.267051 | 0.289086 |
| M | 64 | 0.315082 | 0.289203 | 0.182953 | 0.257807 |

U, offset uniform; L, oblique low-density; M, asymmetric mixed. Values are model probabilities, not percentages or calibrated clinical risks. Each row uses its slope-specific optimized static anchor. At every nonzero slope, the minimizing terminal plans for scenarios 1-4 were the distinct archived endpoints endpoint_0 through endpoint_3. Values are shown to six decimal places; full-precision matrices and the extraction script are included in the reproducibility materials.

**Numerical verification and sensitivity scope**

The source modules retain separate numerical checks for static and probe optimization, policy evaluation, and input integration. Full computational reconstruction passed 52 parent-plan tests, 33 probe-generation tests, 26 adaptive-policy tests, and eight result-synthesis tests. These counts describe tested invariants and archived fixtures, not independent experiments. An independent absolute-risk recursion evaluated all 144 saved policies (72 distinct rules), with maximum scenario-gain discrepancy $6.77327 \times 10^{-11}$ expected events and no checkpoint-minimum discrepancy. The adaptive-policy checks comprised 18 full history trees and 18 state-flow linear programs; the probe-generation checks supplied 12 of each. These checks were executed as part of the full computational reconstruction.

Fifty-five action-label differences between two archived kernels occurred only at terminal states with no remaining uses and had no value effect. No nonterminal action difference was reported in that kernel comparison. Original failed numerical attempts and their corrections are retained rather than replaced by success-only records.

Independent recursion verifies a numerical quantity through another formulation; changing the integration operator assesses sensitivity within the same dose surrogate. Neither establishes a calibrated clinical risk, independent radiation transport, or machine deliverability. The reporting threshold of $10^{-8}$ expected events is an engineering display threshold, not statistical or clinical significance.

**Computational reproducibility and materials**

The reproducibility materials comprise source code, fixed configurations, geometry definitions, parent fluences, score and risk matrices, saved decision rules, numerical summaries, figure data, and verification records. Companion diagnostic files contain the full-precision static risks, probe-eligibility checks, checkpoint-value comparisons, and conditional acquisition counts reported here. The materials contain no patient-level data.

The workflow supports three distinct tasks: reconstructing reported tables and figures from saved outputs; reevaluating frozen policies and descriptive diagnostics; and rebuilding the model from geometry through dose calculation, plan optimization, probe generation, and adaptive-policy evaluation. Full reconstruction passes newly generated outputs to downstream calculations. The large planning-matrix cache is regenerated from source rather than duplicated in the compressed materials.

Full reconstruction was completed by executing the component commands in dependency order with fixed configurations. The recorded environment and thread settings are provided with the code, together with entry-point commands, input-output dependencies, and file checksums. Reproduction uses an initially empty output directory and preserves archived inputs. The README distinguishes the executed component workflow from the packaged launcher, whose directory-layout and dependency checks do not constitute an additional full reconstruction.

**Scope of inference**

The benchmark concerns a finite response family, a specified candidate hull, a fixed horizon and prior, a common terminal menu, an acquisition-risk allowance, and a restricted switching architecture. The four response laws are supplied to the algorithms; the unknown is a common scenario label rather than individual patient radiosensitivity. Immediate, complete outcomes are assumed. The results do not establish performance under unrestricted active control, delayed toxicity, unmodeled response laws, or patient-level heterogeneity.

These exploratory computations are not a confirmatory trial or clinical validation. Negative or unstable increments do not exclude useful informative interventions under different, independently supported models.

**Implementation sequence and interpretation checks**

For each geometry, retain the parent fluences and all score operators with their original normalization. For each response slope, reconstruct the full plan-risk matrices before evaluating any strategy. Optimize the static anchor under the design prior, generate candidate probes under the two-operator acquisition constraints, and form a shared finite terminal menu. The probe-generation objective can be nonsmooth when the minimum-risk terminal plan changes, so each stored local solution must retain its initial point, solver status, feasibility check, and objective value. A completed local solve is not a full-space optimality certificate.

For fixed sampling, evaluate the binomial likelihood for every possible event count, select the minimum posterior-risk terminal plan, and sum the joint-probability-weighted loss. Include both n acquisition uses and H-n terminal uses. At n=0, the procedure must recover a no-acquisition choice from the same terminal menu; at n=H, the terminal contribution is zero. The static anchor is an admissible acquisition choice, so a selected fixed-active design may equal a fixed-passive design. These are boundary checks, not new study conditions.

For passive stopping, evaluate the terminal value and then work backward over all count states. Store the action and selected terminal plan at each state, using the archived tie convention. At a specified active checkpoint, evaluate each eligible single-probe subproblem under the current posterior; compare those values with the already available passive continuation. Propagate the checkpoint value backward through the static prefix. The full active action map must be evaluated separately conditional on each true scenario, rather than treating the optimized prior average as a clinical validation result.

To evaluate a frozen rule under another integration operator, preserve the saved action map and replace both the observation probabilities and the risks of eventual treatments. Changing only the latter would describe a different counterfactual experiment. For a prior-weight sensitivity check, evaluate the same world-conditional gains with the new weights; do not silently optimize a new strategy. For an acquisition-menu ablation, keep the complete terminal menu fixed and change only which fluences may generate observations. A generated plan can be excluded from acquisition without being withheld as a treatment option.

### Comparator decomposition and units

Let $L_{hist}$ be the expected full-horizon loss of a historical reference, $L_{stat}$ the loss of the reoptimized static anchor, $L_{pass}$ the loss of a specified passive learning rule, and $L_{act}$ the loss of a specified active rule. Algebraically, $L_{hist} - L_{act}$ equals $(L_{hist} - L_{stat}) + (L_{stat} - L_{pass}) + (L_{pass} - L_{act})$. This identity is an accounting device, not an additional fitted result or a proof that every term is positive. It shows why an improvement relative to a historical plan alone does not identify an active-information contribution.

In the reported tables, gains use a common reoptimized static anchor within each geometry and slope. The one-switch increment is $L_{pass} - L_{act}$, not the entire active gain and not the improvement over the original physical normalization plan. Counts are expectations across the 2,816-use horizon. Multiplication of a probability difference by 100 gives percentage points; multiplication by the number of uses gives expected event count. Neither operation turns the synthetic response into an empirical clinical rate.

The best static plan depends on the prior; a posterior-specific continuous-hull optimum need not lie in the shared finite terminal menu. A negative frozen-policy increment under another operator or prior does not refute optimization of the nominal input problem, but shows that nominal positivity is not input-independent superiority.